\documentclass[10pt, conference, letterpaper]{IEEEtran}

\IEEEoverridecommandlockouts
\usepackage{cite}
\usepackage{amsmath,amssymb,amsfonts}
\usepackage{algorithm}          
\usepackage[noend]{algpseudocode}  
\usepackage{algpseudocode}
\usepackage{graphicx}
\usepackage{textcomp}
\usepackage{multirow}
\usepackage{hyphenat}
\usepackage{booktabs}
\usepackage[flushleft]{threeparttable}
\usepackage{array}
\usepackage{makecell}
\usepackage[normalem]{ulem}
\usepackage{tabularx}
\usepackage{marvosym}   

\usepackage{xcolor}
\usepackage[colorlinks=true, linkcolor=black, citecolor=black, urlcolor=black]{hyperref}

\def\BibTeX{{\rm B\kern-.05em{\sc i\kern-.025em b}\kern-.08em
    T\kern-.1667em\lower.7ex\hbox{E}\kern-.125emX}}

\begin{document}

\title{STAR: Semantic-Traffic Alignment and Retrieval for Zero-Shot HTTPS Website Fingerprinting
}


\author{
Yifei~Cheng$^{1,2}$,
Yujia~Zhu$^{1,2}$(\Letter),
Baiyang~Li$^{1,2}$,
Xinhao~Deng$^{3}$,
Yitong~Cai$^{1,2}$,
Yaochen~Ren$^{1,2}$,
Qingyun~Liu$^{1,2}$
\\
$^{1}$Institute of Information Engineering, Chinese Academy of Sciences, Beijing, China \\
$^{2}$School of Cyber Security, University of Chinese Academy of Sciences, Beijing, China \\
$^{3}$Institute for Network Sciences and Cyberspace, Tsinghua University, Beijing, China \\
Email: \{chengyifei, zhuyujia, libaiyang, caiyitong, renyaochen, liuqingyun\}@iie.ac.cn, dengxinhao@tsinghua.edu.cn
}

\maketitle



\begin{abstract}

Modern HTTPS mechanisms such as Encrypted Client Hello (ECH) and encrypted DNS improve privacy but remain vulnerable to website fingerprinting (WF) attacks, where adversaries infer visited sites from encrypted traffic patterns. Existing WF methods rely on supervised learning with site-specific labeled traces, which limits scalability and fails to handle previously unseen websites. We address these limitations by reformulating WF as a zero-shot cross-modal retrieval problem and introducing STAR. STAR learns a joint embedding space for encrypted traffic traces and crawl-time logic profiles using a dual-encoder architecture. Trained on 150K automatically collected traffic–logic pairs with contrastive and consistency objectives and structure-aware augmentation, STAR retrieves the most semantically aligned profile for a trace without requiring target-side traffic during training. Experiments on 1,600 unseen websites show that STAR achieves 87.9\% top-1 accuracy and 0.963 AUC in open-world detection, outperforming supervised and few-shot baselines. Adding an Adapter with only four labeled traces per site further boosts top-5 accuracy to 98.8\%. Our analysis reveals intrinsic semantic–traffic alignment in modern web protocols, identifying semantic leakage as the dominant privacy risk in encrypted HTTPS traffic. We release STAR’s datasets and code to support reproducibility and future research\footnote{\href{https://github.com/2654400439/STAR-Website-Fingerprinting}{https://github.com/2654400439/STAR-Website-Fingerprinting}}.

\end{abstract}


\section{Introduction}

As modern HTTPS evolves, traditional protocol-visible identifiers such as Server Name Indication (SNI) and DNS queries are increasingly concealed by mechanisms like Encrypted Client Hello (ECH) \cite{ietf-tls-esni-25} and encrypted DNS \cite{doh_rfc}. This shift limits the effectiveness of conventional web inference techniques that rely on such metadata \cite{ech_2025}. However, even when both payloads and headers are fully encrypted, traffic traces still reveal structural patterns—such as packet sizes, timing, and burst behaviors—that reflect the underlying resource structure of websites \cite{survey_shen}. Website fingerprinting (WF) approaches \cite{hintz_first_wf, k_fp, df, hw, deng_sp, holmes_ccs24} exploit these residual features to infer the site being visited, without requiring access to any plaintext identifiers. In this context, WF has emerged as one of the few remaining passive techniques for web-level inference under full encryption.

Existing WF approaches, however, face fundamental limitations that hinder their scalability and practicality for real-world deployment.  Specifically: (i) Traffic drift. Website content evolves dynamically over time~\cite{concept_drift_analysis}, necessitating frequent recollection of labeled traffic data and retraining of models; (ii) Limited recognition capability. Current supervised learning–based approaches can only identify previously known websites, lacking the ability to generalize to newly emerging sites. These challenges significantly restrict the applicability of WF in operational settings.

To address these limitations, we introduce a novel approach that jointly exploits \textbf{traffic modality} features and \textbf{logical modality} features to enable scalable and generalizable WF against previously unseen websites. Logical modality features (e.g., URI lengths, response sizes, and protocol versions) can be automatically extracted through large-scale web crawling, capturing resource-level attributes that describe a website’s semantic structure. By mapping both traffic modality features and logical modality features into a shared embedding space, we construct a large-scale website fingerprint database grounded in logical representations. Consequently, the task of identifying a website from unseen traffic can be reformulated as a cross-modal retrieval problem, wherein traffic modality features are matched to the most semantically relevant logical modality features stored in the fingerprint database.


We instantiate this formulation through \textbf{STAR} (Semantic–Traffic Alignment and Retrieval), a dual-encoder architecture that jointly embeds logic and traffic modalities into a unified latent space. STAR is trained on over 150K automatically collected logic–traffic pairs using a contrastive learning objective, with additional auxiliary losses to improve intra-class consistency and discriminability. To further enhance robustness against website evolution, we introduce a structure-aware data augmentation mechanism that perturbs both modalities in a semantically consistent manner. During inference, STAR retrieves the most semantically aligned logic profile for an encrypted traffic sample, using cosine similarity in the shared embedding space. This design enables zero-shot classification of encrypted traces with no prior access to traffic from target websites.

Beyond the system design, we also conduct a systematic investigation into \textbf{why semantic–traffic alignment is possible}. We identify three core alignment anchors—on the request side, response side, and transport protocol—each capturing a consistent mapping between traffic features and high-level website structures (§\ref{subsec:key_observ}). These anchors stem from the inherent design of modern web protocols (e.g., header compression, layered transport) and serve as empirical foundations for learning cross-modal associations, further supported by modality-level analyses of discriminability, stability, and cross-modal correlation (§\ref{subsec_modal_selection}). Together, these findings not only validate the design rationale behind STAR, but also provide foundational evidence that cross-modal modeling is both feasible and effective for fingerprinting encrypted web traffic.

In summary, our contributions are as follows:

\begin{itemize}
    \item We formalize \textbf{zero-shot website fingerprinting} under HTTPS as a cross-modal retrieval task, removing the need for per-site traffic collection and supporting generalization to unseen websites.

    \item We present \textbf{STAR}, the \textit{first} dual-modality system that aligns crawl-time semantic logic with encrypted traffic traces through contrastive learning and structure-aware augmentation.

    \item We provide \textbf{empirical and statistical evidence} that semantic–traffic alignment is structurally grounded, revealing significant correlations across multiple alignment anchors.

    \item We perform \textbf{extensive closed- and open-world evaluations}, showing that zero-shot STAR achieves 87.9\% accuracy over 1,600 unseen sites and a 0.963 AUC in an open-world test with millions of distractors, outperforming state-of-the-art supervised and few-shot baselines.

    \item We release the \textbf{STAR-200K dataset and source code} to facilitate future research on semantic inference under encrypted protocols \cite{star_code}.
    
\end{itemize}

These results highlight the feasibility of zero-shot traffic-based identification and demonstrate that \textbf{semantic leakage}, rather than header visibility, now constitutes the principal privacy risk in the encrypted web.


\section{Background and Threat Model}

\subsection{Website Fingerprinting}



Website fingerprinting (WF) infers a user’s visited website by analyzing features of encrypted traffic—such as packet lengths, directions, and timing patterns. Introduced formally by Hintz~\cite{hintz_first_wf}, early WF methods used handcrafted features and classical classifiers~\cite{k_fp, cumul}. The rise of deep learning significantly boosted performance: models like Deep Fingerprinting (DF)~\cite{df} achieved high closed-world accuracy via CNNs, and later work explored GNNs ~\cite{tifs25_http1_wf}, Transformers~\cite{infocom25_wf}, and diffusion models~\cite{diffusion_wf}.


Recent work has revisited WF under mainstream HTTPS, revealing that even minimal protocol interactions can leak identifying patterns. For example, Cebere et al.\cite{raid_wf} analyzed leakage across TLS stages; Gao et al.\cite{tifs25_http1_wf} constructed resource graphs to model site structure; Cheng et al.\cite{hw} showed that HTTP version features form unique site-level sequences; Shen et al.~\cite{www23_tls_wf} demonstrated the use of prior fingerprints to filter obfuscation flows. Other works revealed structural leakage in HTTP/3 and DoH traffic~\cite{siby_quic_wf, doh_wf}.



To reduce the reliance on large labeled datasets, recent work has explored data-efficient strategies under \textit{low-data regimes}. Few-shot approaches use contrastive learning to pretrain DF-style encoders~\cite{tf, NetCLR} or apply kNN over application-layer features~\cite{hw}; generative methods augment training data with synthetic traces~\cite{gan_wf}. Some studies extend to zero-shot scenarios across network conditions (e.g., VPN changes)~\cite{infocom25_wf, gene_wf}, but still assume access to traffic from target websites.

This limitation motivates our proposed cross-modal approach, eliminating the necessity for target-site traffic collection entirely.

\subsection{Threat Model and Problem Definition}
We consider a standard passive adversary in the website fingerprinting (WF) setting \cite{df, hw, www23_tls_wf}. The attacker resides on the network path between the user and the web server—such as an ISP or router—and is able to observe encrypted traffic but cannot modify, delay, inject, or decrypt any packets. The attacker's goal is to determine whether the user is visiting a monitored website and, if so, identify which one.

Unlike traditional WF approaches that formulate this task as a site-specific classification problem, we adopt a new threat model where the attacker performs \textbf{cross-modal retrieval} between encrypted traffic traces and semantic website representations. This allows the attacker to recognize previously unseen websites based on semantic-traffic alignment, without requiring labeled traffic samples for each monitored site.

\section{Core Observations and Cross-Modal Alignment Hypothesis}
\label{sec:alignment}

\begin{figure}[tp]
    \centering
    \includegraphics[width=\linewidth]{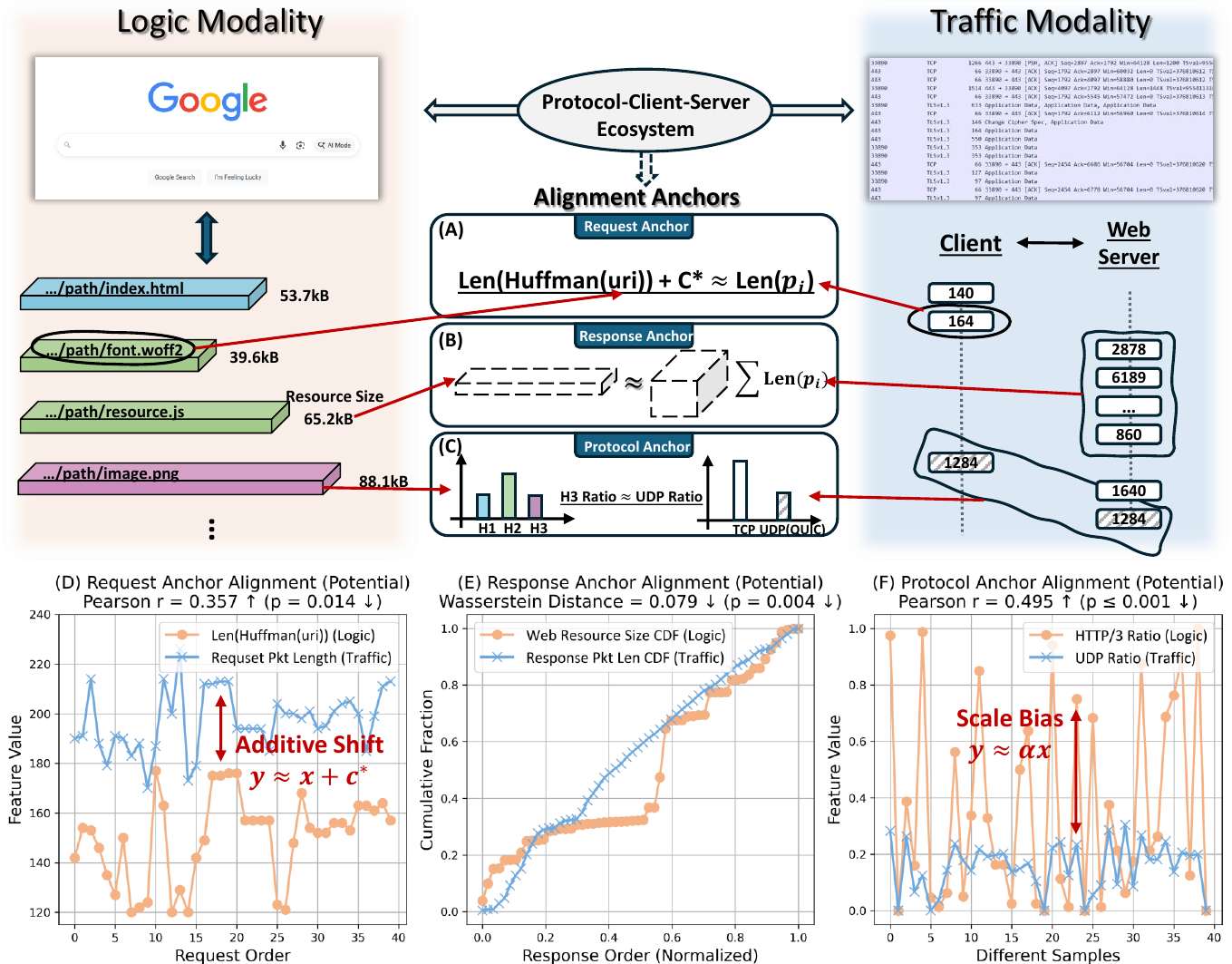}
    \caption{
        Overview of cross-modality alignment anchors in our setting. 
        (A–C) illustrate three hypothesized alignment anchors between website semantic logic (left) and encrypted traffic behavior (right): request-side, response-side, and transport protocol. 
        (D–F) present empirical support for each anchor via Pearson correlation or Wasserstein distance on representative samples.
    }
    \label{fig:key_observation}
\end{figure}

\subsection{Cross-Modal Design Principles}
\label{para:modal_design_p}
Unlike typical cross-modal learning tasks \cite{CLIP} that align well-structured modalities such as natural language and images, our setting involves the alignment between encrypted traffic traces and abstracted semantic representations of websites. This \textbf{non-traditional modality pairing} introduces new challenges, particularly in the construction of effective input representations.

To guide the design of both traffic and logic modalities, we summarize three key principles that effective cross-modal representations should satisfy, inspired by prior work in fingerprinting and multi-modal retrieval:

\begin{itemize}
    \item \textbf{P1: Discriminability (intra-modality)}
    
    Each modality should encode features that allow websites to be distinguished from one another, enabling the model to separate classes in both semantic and traffic spaces.
    
    \item \textbf{P2: Stability (intra-modality)}
    
    The modality representations should remain relatively consistent across repeated visits to the same site under similar conditions. High intra-class consistency is critical for generalization.
    
    \item \textbf{P3: Alignability (cross-modality)}
    
   There should exist identifiable structural relationships between the modalities, which we refer to as \textbf{alignment anchors}. These anchors serve as a learnable bridge that enables the model to connect encrypted traffic behavior with semantic site characteristics.
\end{itemize}


These principles form the foundation for our modality design choices and motivate the structural observations presented in the following sections.

\subsection{Core Alignment Observations}
\label{subsec:key_observ}
We define a \textbf{cross-modal alignment anchor} as a feature or structure that exhibits semantic correspondence and measurable correlation between the logic and traffic modalities. In our setting, we identify three such alignment anchors—on the request side, response side, and transport protocol—each rooted in the design of modern web communication ecosystems (see Fig. \ref{fig:key_observation} A–C).

\subsubsection{Request Anchor}

Our key observation is that the \textbf{length of HTTPS request packets} is linearly related to the \textbf{Huffman-encoded length of the resource URI}. This arises from protocol-level optimizations in HTTP/2 and HTTP/3, which employ static/dynamic header compression \cite{rfc7541_http2_compression}. Most headers (e.g., User-Agent, Cookie) are replaced with compact indices, leaving the URI as the dominant uncompressed field, further compressed by a public Huffman table. Thus, request packet length can be approximated as:
\begin{equation}
Len(p_i) \approx \mathrm{Len}(\mathrm{Huffman}(uri_i)) + C \times H
\end{equation}
where \(p_i\) is the packet length and \(H\) is the number of compressed headers. This alignment is visualized in Fig.\ref{fig:key_observation} A and supported in Fig.\ref{fig:key_observation} D.

\subsubsection{Response Anchor}
Response packets convey web content, and their cumulative size naturally reflects the sum of individual resource sizes \cite{FineWP}. We observe that:
\begin{equation}
\sum Len(p_i)^{(\mathrm{resp})} \approx \mathrm{Size}(resource_i)
\end{equation}
enabling logic-to-traffic comparison of response behavior (Fig. \ref{fig:key_observation}B, E).

\begin{table}[t]
\centering
\caption{Evaluation of alignment anchors across top-1000 websites.}
\label{tab:alignment_stats}
\begin{tabularx}{\linewidth}{@{} l l c c c @{}}
\toprule
\textbf{Anchor} & \textbf{Metric} & \textbf{Mean Value} ↑ & \textbf{p-value} ↓ & \textbf{Sig. (\%)} ↑ \\
\midrule
Request  & Pearson $r$            & 0.3114 ± 0.0730   & 0.0290  & 86 \\
Response & $1 -$ Wasserstein & 0.9109 ± 0.0278   & 0.0124  & 96 \\
Protocol & Pearson $r$            & 0.5607 ± 0.1019   & 0.0017  & 100 \\
\bottomrule
\end{tabularx}
\end{table}

\subsubsection{Protocol Anchor}
HTTP/3 operates over QUIC/UDP, creating observable transport-layer patterns \cite{hw}. We compare the \textbf{UDP traffic ratio} with the \textbf{server-side HTTP/3 usage ratio}, forming a protocol anchor:
\begin{equation}
\mathrm{UDP\ Ratio} \approx \mathrm{HTTP/3\ Usage\ Ratio}
\end{equation}
This structural similarity enables indirect inference of protocol usage (Fig. \ref{fig:key_observation}C, F).


To quantify these alignments, we perform statistical \textbf{hypothesis testing} on paired samples. For each anchor, we extract matched feature sequences from both modalities, apply normalization when necessary, and measure alignment using Pearson correlation or Wasserstein distance. Significance is evaluated via \textbf{permutation testing}. Aggregated results in Table~\ref{tab:alignment_stats} confirm statistically significant alignment across all three anchors, motivating the modality representations used in our framework (§\ref{sec_method}).

\section{Methodology}
\label{sec_method}
\begin{figure}[tp]
    \centering
    \includegraphics[width=\linewidth]{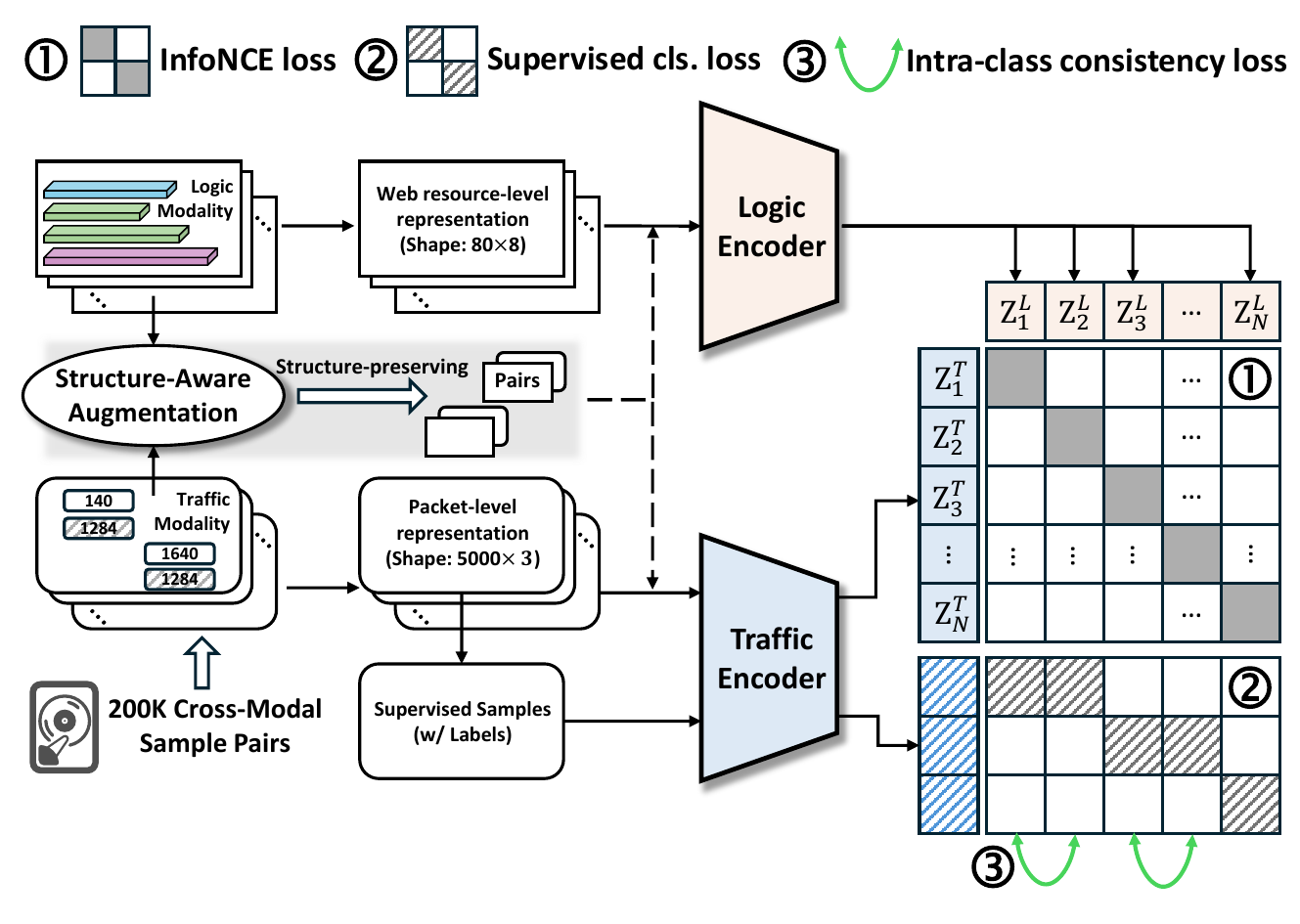}
    \caption{
        \textbf{Training‑stage framework of STAR.}
Structure‑aware logic–traffic sample pairs and labeled traffic samples are passed through the Logic Encoder and Traffic Encoder, whose weights are jointly learned so that (1) paired embeddings align via contrastive loss, (2) traffic embeddings support supervised classification, and (3) same‑class traffic embeddings remain consistent.
    }
    \label{fig:framework}
\end{figure}

This section presents the STAR framework for semantic–traffic alignment in website fingerprinting. As shown in Fig.~\ref{fig:framework}, STAR maps paired inputs from two heterogeneous modalities—website logic and encrypted traffic—into a shared embedding space for unified retrieval and classification.


We first define modality-specific input representations (§\ref{subsec_representation}), then design a dual-encoder architecture to embed them (§\ref{sec:encoder}). The encoders are jointly trained with contrastive and auxiliary losses to promote semantic alignment (§\ref{subsec_loss}). To enhance generalization, we introduce structure-aware data augmentation (§\ref{sec:augmentation}). The learned encoders support flexible downstream usage, including zero-shot retrieval and few-shot adaptation (see Fig. \ref{fig:test_framework}).

\subsection{Framework Overview}
\label{sec:overview}

\begin{figure}[tp]
    \centering
    \includegraphics[width=\linewidth]{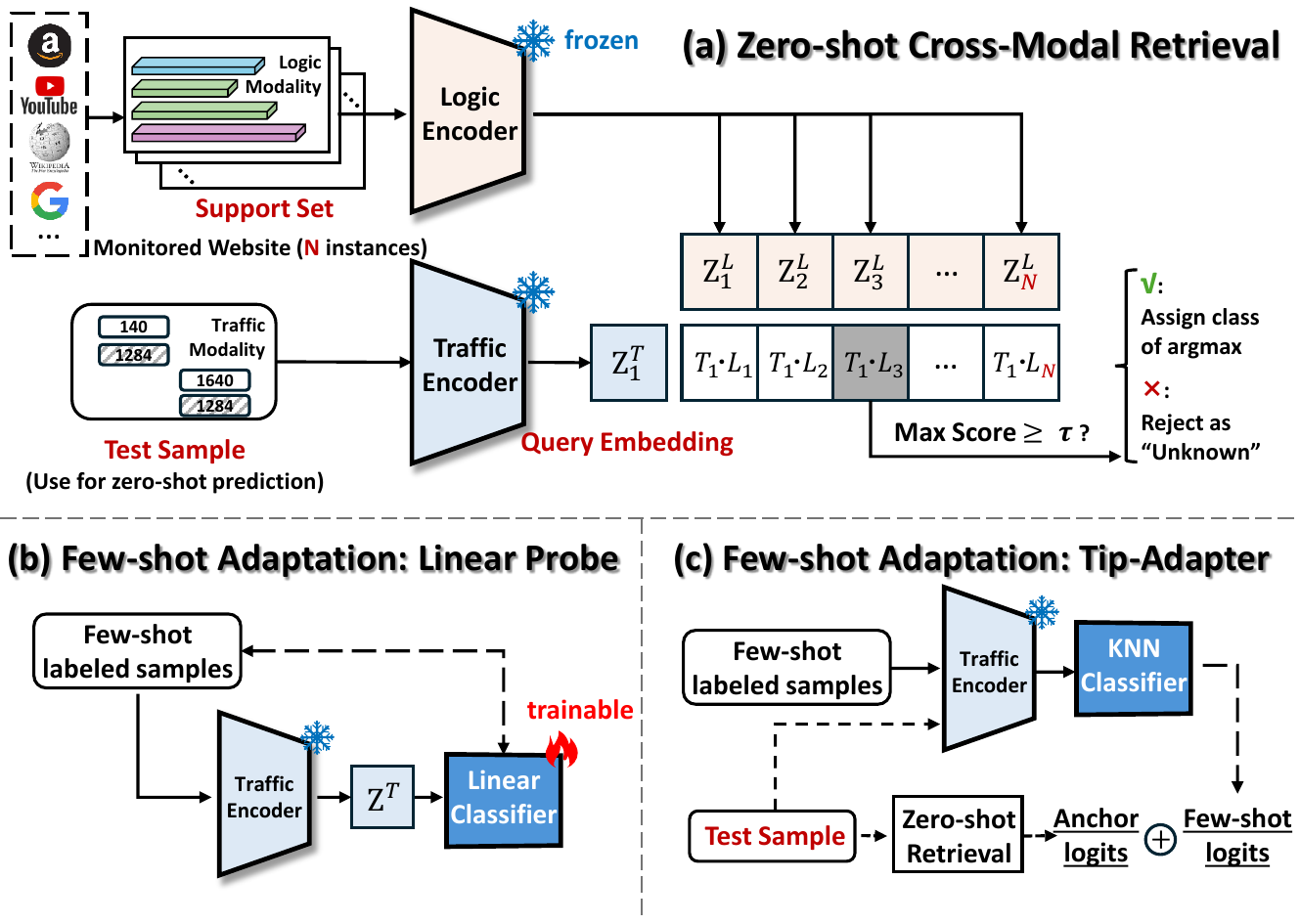}
    \caption{
    \textbf{Inference-stage framework of STAR.}
    (a) \textbf{Zero-Shot Retrieval}: encode a test trace and match it against gallery logic embeddings; assign the top class if the similarity exceeds a threshold, otherwise reject as “Unknown.”
    (b) \textbf{Few-Shot Linear Probe}: train a linear classifier on few-shot traffic embeddings with the encoder frozen.
    (c) \textbf{Few-Shot Tip-Adapter}: fuse anchor-based logits from logic retrieval with k-NN logits from a few-shot traffic memory for final prediction.
    }

    \label{fig:test_framework}
\end{figure}

Our STAR framework is built on a cross-modal dual-encoder design, as illustrated in Fig.~\ref{fig:framework} and Fig.~\ref{fig:test_framework}. It takes as input two modalities for each website access:  
(i) a \textit{logic modality}, which encodes semantic web resource structures—such as resource uri lengths, sizes, and protocol behaviors—and  
(ii) a \textit{traffic modality}, which captures encrypted packet-level features during access. These paired inputs are processed by separate encoders and projected into a shared embedding space. To enforce semantic alignment, we apply an InfoNCE-based contrastive loss, while auxiliary classification and consistency losses promote inter-class separability and intra-class coherence.

During training (Fig.~\ref{fig:framework}), STAR is optimized on large-scale cross-modal sample pairs collected via automated crawling and traffic capture, further expanded through structure-aware data augmentation. The learning objectives jointly update both encoders to align paired embeddings, classify traffic samples, and cluster instances from the same class.


At inference time (Fig.~\ref{fig:test_framework}), the trained encoders support multiple downstream scenarios. For \textit{zero-shot classification}, a test trace is encoded by the traffic encoder and compared—via cosine similarity in the shared embedding space—against a gallery of logic-side prototypes pre-computed from crawl-time profiles. The top-matched class is returned if the similarity exceeds a decision threshold; otherwise, the input is rejected as unmonitored. For \textit{few-shot adaptation}, STAR integrates with plug-and-play strategies such as linear probing \cite{CLIP} or Tip-Adapter-style fusion \cite{tipadapter}, both operating over frozen encoders 
\footnote{
Detailed implementation-level descriptions of inference procedures are provided in an online technical appendix:
\url{https://github.com/2654400439/STAR-Website-Fingerprinting/blob/main/docs/Technical Appendix/STAR_Technical_Appendix.pdf}.}.
This retrieval-based formulation enables scalable, flexible deployment in open-world scenarios without requiring retraining or per-target traffic collection.

\subsection{Modality Representation Construction}
\label{subsec_representation}

To enable reliable cross-modal alignment, we design compact yet expressive representations for encrypted traffic and site logic. Each is structured as a fixed-length sequence with features selected to reflect the core alignment anchors identified in §\ref{subsec:key_observ}, while maintaining generalizability and model efficiency.

\subsubsection{Traffic Modality}


We represent encrypted traffic traces as a sequence of packet-level features, defining a feature matrix $T \in \mathbb{R}^{5000 \times 3}$, where each row $f_i^{(T)} = \left[ \mathrm{dir}(p_i),\; v_i,\; s_i \right]$ corresponds to a packet $P_i$:
\begin{itemize}
    \item $\mathrm{dir}(p_i)$ is the \textit{directional packet length}, where client-to-server packets are positive and server-to-client packets are negative. This single value reflects both request and response behaviors, preserving alignment signals from both ends.
    \item $v_i \in \{1,2,3\}$ is the \textit{inferred HTTP version}. Inspired by \cite{hw}, we heuristically assign this per-packet label based on transport-layer characteristics: UDP packets are labeled HTTP/3; TCP packets are marked HTTP/2 if two consecutive packets begin with TLS content-type \texttt{0x17}\footnote{0x17 is the TLS \texttt{content\_type} value indicating \textit{application data} \cite{rfc_tls13}}, otherwise HTTP/1.1.
    \item $s_i \in \mathbb{Z}^+$ is the \textit{flow index}, indicating the bidirectional connection to which the packet belongs, enabling coarse-grained structural grouping within traces.
\end{itemize}


\subsubsection{Logic Modality}

The logic modality encodes a website’s resource-level structure as a semantic matrix $L \in \mathbb{R}^{80 \times 8}$, where each row $f_j^{(L)}$ corresponds to a web resource as observed during page load. These resource vectors capture the website’s high-level semantics and are extracted from browser developer logs \cite{chromelog} via automated scripts.

We group the eight features into three semantic categories:
\begin{itemize}
    \item \textbf{Identifier length indicators}: Huffman-encoded and raw URI lengths provide a compact representation of the resource path size and align with the request-side packet lengths in the traffic modality.
    \item \textbf{Content indicators}: Response size and header length describe the volume of returned data per resource, enabling alignment with response-side traffic features.
    \item \textbf{Protocol-level context}: HTTP version, alternative service flag (for HTTP/3 support), MIME type category, and server IP index encode protocol semantics and content-type variability across resources, supporting protocol-aware mapping and resource grouping.
\end{itemize}

\subsubsection{Normalization \& Encoding}

All inputs are formatted as fixed-length matrices: traffic traces are truncated or zero-padded to 5000 packets, and logic traces to 80 resources. Continuous features (e.g., lengths and sizes) are log-scaled, categorical fields (e.g., MIME type, stream index) are embedded via learnable vectors, and Boolean values are represented as binary integers.

These design choices ensure that the learned representations remain compact, semantically meaningful, and structurally aligned across modalities.

\subsection{Dual-Encoder Architecture}
\label{sec:encoder}
To bridge the modality gap between encrypted traffic traces and website logic structures, we adopt a dual-encoder architecture to project each modality into a shared embedding space. This architecture is inspired by the CLIP \cite{CLIP} paradigm, where modality-specific encoders are used to preserve intra-modality semantics while enabling cross-modal alignment via \textbf{contrastive training}.

\paragraph{Traffic Encoder}
For the traffic modality, we build upon the DFNet \cite{df} backbone, a deep convolutional network widely adopted in website fingerprinting literature due to its strong discriminative capacity under encrypted traffic. Given a packet-level input matrix $\mathbf{T} \in \mathbb{R}^{5000 \times 3}$, we replace the original 1D convolutional layers with three-channel convolutions to accommodate the 3-dimensional packet features. We remove the classification head of DFNet and preserve the penultimate hidden representation as the traffic embedding. A subsequent projection head $f_T$ maps the encoder output to a normalized embedding:
\begin{equation}
\mathbf{z}_i^T = \frac{f_T\left( \text{DFEnc}(\mathbf{T}_i) \right)}{ \left| f_T\left( \text{DFEnc}(\mathbf{T}_i) \right) \right|_2 }
\end{equation}

\paragraph{Logic Encoder}
For the logic modality, we employ a Transformer encoder \cite{transformer} to effectively process structured sequences of web resources. Given a resource-level input matrix $\mathbf{L} \in \mathbb{R}^{80 \times 8}$, the encoder utilizes multi-head self-attention to capture feature-wise and resource-wise dependencies, allowing the model to learn hierarchical importance among resources. The output representations are aggregated via masked average pooling, followed by a projection head $f_L$ that yields the normalized logic embedding:
\begin{equation}
\mathbf{z}_i^L = \frac{f_L\left( \text{TransEnc}(\mathbf{L}_i) \right)}{ \left| f_L\left( \text{TransEnc}(\mathbf{L}_i) \right) \right|_2 }
\end{equation}

Each embedding $\mathbf{z}_i^L$ or $\mathbf{z}_i^T$ resides in a shared latent space $\mathbb{R}^{d}$ (we set $d=256$), which serves as the basis for cross-modal contrastive alignment.

\subsection{Cross-Modal Contrastive Training}
\label{subsec_loss}
To align the logic and traffic modalities, we adopt a multi-objective training strategy centered on InfoNCE loss \cite{infonce} and supplemented by auxiliary supervision. The goal is to ensure that matched logic-traffic pairs are closer in the embedding space than mismatched pairs.

\subsubsection{InfoNCE Loss for Cross-Modal Alignment}
We leverage the standard contrastive loss over a batch of $N$ paired samples. For each traffic embedding $\mathbf{z}_i^T$ and its corresponding logic embedding $\mathbf{z}_i^L$, the InfoNCE objective encourages the inner product $\langle \mathbf{z}i^T, \mathbf{z}i^L \rangle$ to be higher than that of any non-matching pair. The loss is given by:
\begin{equation}
\mathcal{L}_{\text{InfoNCE}} = - \sum_{i=1}^{N} \log \frac{ \exp\left( \langle \mathbf{z}_i^T, \mathbf{z}i^L \rangle / \tau \right) }{ \sum_{j=1}^{N} \exp\left( \langle \mathbf{z}_i^T, \mathbf{z}_j^L \rangle / \tau \right) }
\end{equation}
where $\tau$ is a temperature hyperparameter. Unlike conventional supervised contrastive learning, all negatives in the denominator are guaranteed to be true negatives due to the use of large-scale unlabeled pairs across diverse websites (each pair from a distinct site, see §\ref{subsec_dataset}), preventing semantic ambiguity.

\begin{algorithm}[t]
\caption{Structure-Aware Cross-Modal Augmentation}
\label{alg:augmentation}
\begin{algorithmic}[1]
\Require Logic modality $\mathbf{R} = \{r_1, r_2, \dots, r_n\}$ with IP tags; Traffic modality $\mathbf{P} = \{p_1, p_2, \dots, p_m\}$ with IP tags
\Ensure Augmented pair $(\hat{\mathbf{R}}, \hat{\mathbf{P}})$

\State Group $\mathbf{R}$ by server IP: $\mathcal{S} \gets \{s_1, \dots, s_k\}$ with resource groups $\mathcal{G}(s_i)$
\State Compute IP selection scores: $\omega(s_i) \gets 1 - \frac{|\mathcal{G}(s_i)|}{|\mathbf{R}|}$

\State Sample deletion threshold $T \sim \mathcal{N}(\mu=0.3, \sigma=0.1) \cdot |\mathbf{R}|$
\State Initialize $\mathcal{S}_{\text{del}} \gets \emptyset$, $\hat{\mathbf{R}} \gets \mathbf{R}$, $\hat{\mathbf{P}} \gets \mathbf{P}$

\While {total deleted resources $< T$}
    \State Sample IP $s \sim \omega(s)$ with weighted probability
    \If {$s \notin \mathcal{S}_{\text{del}}$}
        \State Remove $\mathcal{G}(s)$ from $\hat{\mathbf{R}}$; remove packets with IP $s$ from $\hat{\mathbf{P}}$
        \State $\mathcal{S}_{\text{del}} \gets \mathcal{S}_{\text{del}} \cup \{s\}$
    \EndIf
\EndWhile

\State \Return $(\hat{\mathbf{R}}, \hat{\mathbf{P}})$
\end{algorithmic}
\end{algorithm}

\subsubsection{Supervised Contrastive Loss for Discrimination}
To further enhance class-discriminative capacity in the traffic modality, we incorporate a supervised contrastive loss using labeled fingerprinting datasets. Following \textit{SupCon} \cite{supcon}, the loss encourages embeddings from the same class to be closer, while keeping different classes apart:
\begin{align}
\mathcal{L}_{\text{SupCon}} = - \sum_{i \in \mathcal{I}} \frac{1}{|\mathcal{P}(i)|} 
\sum_{p \in \mathcal{P}(i)} \log 
\frac{ \exp\left( \langle \mathbf{z}_i^T, \mathbf{z}_p^T \rangle / \tau \right) }
{ \sum_{a \in \mathcal{A}(i)} \exp\left( \langle \mathbf{z}_i^T, \mathbf{z}_a^T \rangle / \tau \right) }
\end{align}

where $\mathcal{P}(i)$ denotes the set of positives (same class), and $\mathcal{A}(i)$ the set of all anchors except $i$.

\subsubsection{Consistency Loss for Stability}
Given the inherent instability of encrypted traffic, even within the same class, we introduce an intra-class consistency loss to promote local smoothness among traffic embeddings. Specifically, we minimize pairwise distance among all traffic embeddings with the same class label:
\begin{equation}
\mathcal{L}_{\text{Consistency}} = \sum_{(i,j) \in \mathcal{C}} \left| \mathbf{z}_i^T - \mathbf{z}_j^T \right|_2^2
\end{equation}
where $\mathcal{C}$ is the set of intra-class traffic pairs.

\subsubsection{Final Objective.}
The full training objective combines all three components with weighting coefficients:
\begin{equation}
\label{eq_loss}
\mathcal{L} = \mathcal{L}_{\text{InfoNCE}} + \lambda{\text{sup}} \mathcal{L}_{\text{SupCon}} + \lambda{\text{cons}} \mathcal{L}_{\text{Consistency}}
\end{equation}

This hybrid objective enables us to exploit both large-scale weakly-aligned web pairs and reliable supervised samples to improve alignment quality and generalization.

\subsection{Structure-Aware Cross-Modal Augmentation}
\label{sec:augmentation}



To enhance generalization under site evolution, we introduce a structure-aware augmentation method that perturbs both modalities in a consistent manner. This approach generates realistic logic–traffic sub-pairs while preserving the structural alignment necessary for contrastive training. Unlike traditional view-level augmentations or modality-specific transformations \cite{declip}, our method exploits a shared structural anchor—\textbf{server IP addresses}—that appears in both modalities and governs subsets of web resources and traffic packets.

The augmentation operates by selectively dropping all resources in the logic modality that are associated with a sampled set of server IPs. The corresponding traffic packets linked to the same IPs are then removed from the traffic modality, producing a semantically valid and internally consistent sub-pair. To avoid excessive content removal, IPs are sampled with inverse probability proportional to their resource count, and deletions continue until a stochastic threshold is met. This threshold is drawn from a Gaussian prior to introduce controlled variation. The full procedure is outlined in Algorithm~\ref{alg:augmentation}.

The resulting augmented pairs preserve partial yet coherent cross-modal alignment and are seamlessly integrated into the training set as additional samples for contrastive learning.

\section{Evaluation}

In this section, we conduct comprehensive experiments to evaluate the effectiveness of \textbf{STAR} across both closed-world and open-world settings. We first describe the datasets used in our experiments (§\ref{subsec_dataset}) and introduce competitive baselines (§\ref{subsec_baseline}). We then perform a series of experiments, including modality design analysis (§\ref{subsec_modal_selection}), classification under closed-world (§\ref{subsec_cw}) and open-world (§\ref{subsec_ow}) settings, as well as in-depth ablation and interpretability analysis (§\ref{subsec_ablation}).

\subsection{Datasets}
\label{subsec_dataset}
We utilize two types of datasets in our experiments: (1) a large-scale cross-modal dataset constructed by ourselves, and (2) an existing labeled fingerprinting dataset used for evaluation and auxiliary supervision.

\textbf{(1) Cross-Modal Dataset (STAR-200K).}  
We collect a large-scale dataset of website-level cross-modal samples based on the top 200,000 sites from the Tranco list\footnote{Available at https://tranco-list.eu/list/5XYPN.} \cite{tranco}. Data collection is performed on ten geographically distributed AWS EC2 instances across North America, Europe, and Asia. For each site, we use Selenium-controlled \texttt{Chrome browsers} (selected for its market share $>$60\% \cite{cloudflareRadarBrowser}) to access the homepage, extracting browser logs for logical modality representation. Concurrently, raw traffic is captured using \texttt{tcpdump} to build the encrypted traffic modality. Each site is accessed once; after filtering for failures (e.g., connection issues, CAPTCHAs), we obtain over 170K valid sample pairs. We refer to this dataset as \textbf{STAR-200K}. We use 150K pairs for training STAR, and reserve 20K disjoint pairs for open-world evaluation.

\newcommand{\secondbest}[1]{\smash{\uline{#1}}}
\newcolumntype{C}[1]{>{\centering\arraybackslash}m{#1}}

\begin{table}[t]
\scriptsize
\setlength{\tabcolsep}{3pt}   
\renewcommand{\arraystretch}{1.15}
\caption{Zero-shot classification results and modality properties with different traffic representations.}
\label{tab:modality_comparison}

\begin{tabularx}{\linewidth}{@{}
    C{1.25cm} C{1.35cm} |   
    *{2}{C{1.0cm}} |        
    *{3}{C{1.0cm}}          
    @{}}
\toprule
\multicolumn{2}{c|}{\textbf{Modality Representation}} &
\multicolumn{2}{c|}{\textbf{Task Accuracy}} &
\multicolumn{3}{c}{\textbf{Modality Properties}}\\
\cmidrule(lr){1-2}\cmidrule(lr){3-4}\cmidrule(l){5-7}
\textbf{Logic} & \textbf{Traffic} &
\textbf{Top‑1 (\%)} & \textbf{Top‑5 (\%)} &
\textbf{AMI (P1)} & \textbf{FDR (P2)} & \textbf{dCor (P3)} \\
\midrule
\multirow{7}{*}{Ours}
  & CUMUL~\cite{cumul}         & 36.69  & 62.48  & 0.3539 &  0.3611      &  0.3811      \\
  & Trace~\cite{df}            & \secondbest{52.44}  & \secondbest{80.19}  & 0.2389 & 0.2445 & \textbf{0.6312} \\
  & H123~\cite{hw}     & 50.50  & 76.62  & \textbf{0.6748} & \textbf{1.909}  & 0.4466 \\
  & TAM~\cite{rf}              & 12.09  &  42.03 & 0.4121 & 1.175  & 0.2791 \\
  & WTCM~\cite{countmamba}     & 18.76  &  50.18 & 0.4893 & 1.2088 & 0.2329 \\
  & Ours                       & \textbf{87.87}  & \textbf{96.94}  & \secondbest{0.6228} & \secondbest{1.5744} & \secondbest{0.5906} \\
\bottomrule
\end{tabularx}
\end{table}


\textbf{(2) Labeled Fingerprinting Dataset (H\&W-1600).}  
We use the public dataset from~\cite{hw}, which provides 40 traffic samples for each of 2,240 HTTPS websites across three groups: \texttt{popular}, \texttt{random}, and \texttt{censorship}. We select the \texttt{popular} subset (1,600 websites) for closed-world evaluation. The remaining samples are used as labeled data for supervised training modules (§\ref{subsec_loss}). To prevent data leakage, we ensure all evaluation websites are disjoint from the STAR-200K pretraining and labeled training sets.

\begin{table*}[t]
  \centering
  \caption{Closed-World Website‑Fingerprinting Accuracy (\%)}
  \label{tab:fewshot}
  \renewcommand{\arraystretch}{0.88}
  \scriptsize
  \resizebox{\textwidth}{!}{%
  \begin{tabular}{l *{6}{cc}}
    \toprule
    \multirow{2}{*}{\textbf{Method}} &
      \multicolumn{2}{c}{\textbf{0‑shot}} &
      \multicolumn{2}{c}{\textbf{1‑shot}} &
      \multicolumn{2}{c}{\textbf{2‑shot}} &
      \multicolumn{2}{c}{\textbf{4‑shot}} &
      \multicolumn{2}{c}{\textbf{8‑shot}} &
      \multicolumn{2}{c}{\textbf{16‑shot}} \\
      \cmidrule(lr){2-3}\cmidrule(lr){4-5}\cmidrule(lr){6-7}
      \cmidrule(lr){8-9}\cmidrule(lr){10-11}\cmidrule(l){12-13}
      & top‑1 & top‑5 & top‑1 & top‑5 & top‑1 & top‑5
      & top‑1 & top‑5 & top‑1 & top‑5 & top‑1 & top‑5 \\
    \midrule
    CUMUL \cite{cumul}          & /      & /     & 64.46 & 73.61 & 72.93 & 79.80 & 76.52 & 85.13 & 84.04 & 90.77 & 87.45 & 92.18 \\
    DF \cite{df}             & /      & /     & 17.48 & 37.44 & 51.76 & 76.71 & 73.04 & 90.72 & 85.13 & 96.39 & 91.46 & 98.50 \\
    DF$+$              & /      & /     & 33.10 & 57.18 & 67.67 & 85.92 & 77.31 & 91.75 & 91.13 & 98.36 & \secondbest{95.41} & 99.35 \\
    RF \cite{rf}             & /      & /     & 26.63 & 47.84 & 51.29 & 75.49 & 65.91 & 86.99 & 76.10 & 92.66 & 79.43 & 94.82 \\
    CountMamba \cite{countmamba}     & /      & /     & 47.16 & 66.68 & 68.34 & 85.09 & 90.04 & 97.73 & \secondbest{93.56} & 98.94 & \textbf{95.62} & \textbf{99.62} \\
    \midrule[\heavyrulewidth]  
    H\&W \cite{hw}           & /      & /     & \secondbest{78.70} & 89.08 & 85.42 & 91.96 & 88.01 & 93.70 & 89.02 & 93.98 & 89.51 & 94.12 \\
    NetCLR \cite{NetCLR}         & /      & /     & 36.67 & 58.74 & 55.60 & 77.35 & 74.77 & 91.02 & 87.17 & 96.62 & 92.51 & 98.22 \\
    TF \cite{tf}             & /      & /     & 59.21 & 59.33 & 69.87 & 78.42 & 76.69 & 89.29 & 80.24 & 92.58 & 82.03 & 93.27 \\
    \midrule[\heavyrulewidth]  
    FineWP \cite{FineWP}         & /      & /     & 35.00 & 53.72 & 66.63 & 82.83 & 82.88 & 93.66 & 88.84 & 96.33 & 92.21 & 97.74 \\
    Oscar \cite{Oscar}          & /      & /     & 46.94 & 64.34 & 63.71 & 79.08 & 76.70 & 88.32 & 82.55 & 91.98 & 86.16 & 93.95 \\
    \midrule[\heavyrulewidth]  
    Clustering + Hungarian & 30.04 & /  & / & / & / & / & / & / & / & / & / & / \\
    \midrule
    STAR‑Linear Probe  & \multirow{2}{*}{\textbf{87.87}} & \multirow{2}{*}{\textbf{96.94}} & 74.53 & \secondbest{92.21} & \secondbest{86.63} & \secondbest{97.12} & \secondbest{91.59} & \secondbest{98.78} & \textbf{94.24} & \textbf{99.38} & 95.06 & \secondbest{99.39} \\
    STAR‑Tip Adapter   &       &      & \textbf{88.26} & \textbf{97.20} & \textbf{90.93} & \textbf{98.52} & \textbf{91.92} & \textbf{98.84} & 93.42 & \secondbest{98.95} & 94.11 & 99.09 \\
    \bottomrule
  \end{tabular}}
\end{table*}

\subsection{Baselines}
\label{subsec_baseline}
To demonstrate the effectiveness of \textbf{STAR}, the first \emph{zero-shot} website fingerprinting method without access to target traffic, we compare against representative state-of-the-art baselines from three categories.

\textbf{Standard WF methods} include CUMUL~\cite{cumul}, which uses cumulative packet lengths with an SVM classifier; DF+~\cite{df}, a CNN-based model extended to directional packet lengths for HTTPS settings; RF~\cite{rf}, which utilizes fixed-time aggregation matrices for deep classification; and CountMamba~\cite{countmamba}, which models coarse-grained count matrices using a state space model for robust, early-stage classification.

\textbf{Few-shot methods} include TF~\cite{tf} and NetCLR~\cite{NetCLR}, both of which pretrain DF-based encoders using contrastive learning (NetCLR adds self-supervised tasks), and H\&W~\cite{hw}, which matches application-layer features via KNN.

\textbf{Fine-grained methods} include FineWP~\cite{FineWP}, using statistical features with random forests, and Oscar~\cite{Oscar}, which applies multi-label metric learning for precise web page classification.

All baselines are implemented using official or WFlib~\cite{wflib} code with default settings.

\subsection{Evaluation of Modality Representations}
\label{subsec_modal_selection}
To validate the effectiveness of our cross-modal formulation, we begin with a systematic evaluation of different modality representation choices.

\noindent\textbf{• Experimental Setup.} We adopt the proposed STAR training paradigm and explore its behavior under various traffic modality representations. The logic modality is fixed to our proposed 8-dimensional web resource-level representation, while the traffic modality varies across prior designs in the website fingerprinting literature. For example, we include Trace sequences \cite{df}, flow-level statistical summaries (H123) \cite{hw}, the Traffic Aggregation Matrix (TAM) \cite{rf}, and the Windowed Traffic Counting Matrix (WTCM) \cite{countmamba}. For each modality combination, we perform full model training with our multi-loss objective and structure-aware augmentation, then evaluate zero-shot classification performance on the H\&W-1600 dataset.

\begin{figure}[tp]
    \centering
    \includegraphics[width=\linewidth]{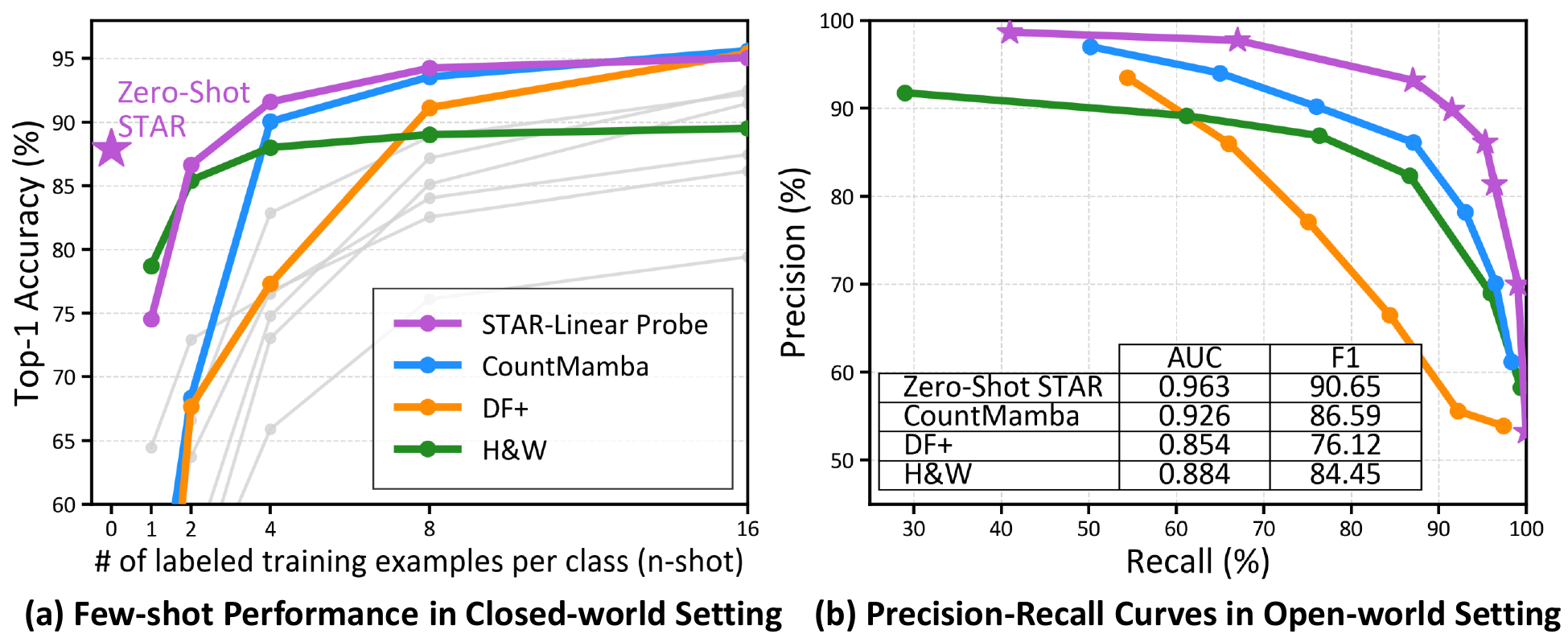}
    \caption{
        \textbf{Closed-world and open-world performance comparison.}
\textbf{(a)} Top-1 accuracy under different n-shot settings in the closed-world scenario. Zero-shot STAR is marked with a purple star, and the top-3 few-shot baselines are color-highlighted; others are shown in grey.
\textbf{(b)} Precision-recall curves in the open-world 4-shot setting, comparing Zero-shot STAR with the top-3 baselines. AUC and best F1 scores are shown in the table.
    }
    \label{fig:cw_ow}
\end{figure}

To complement accuracy metrics, we additionally assess three modality design criteria introduced in §\ref{para:modal_design_p}:

\begin{itemize}
    \item \textbf{Inter-class discriminability (P1)} is quantified via \textbf{Adjusted Mutual Information (AMI)}, which measures how well the traffic embeddings can be clustered into groups that match the true class labels.

    \item \textbf{Intra-class stability (P2)} is estimated by the \textbf{Fisher Discriminant Ratio (FDR)}, comparing between-class and within-class variances of traffic embeddings.
    
    \item \textbf{Cross-modal alignability (P3)} is captured using \textbf{Distance Correlation (dCor)} \cite{dcor} between normalized embeddings ${ \mathbf{z}_i^T }$ and ${ \mathbf{z}_i^L }$. dCor equals zero if and only if two random variables are statistically independent.
\end{itemize}

All statistics are computed on traffic embeddings normalized to zero mean and unit variance. AMI and dCor are scale-invariant, while FDR is already a scale-free ratio.

\noindent\textbf{• Results.} 
As shown in Table~\ref{tab:modality_comparison}, Trace-based representations achieve the highest dCor, benefiting from explicit request-response anchoring that naturally aligns with logic semantics. H123 obtains the best AMI and FDR scores, thanks to its protocol-aware descriptors and flow-level aggregation that enhance class discriminability and intra-class consistency.



However, these statistical strengths do not directly translate into strong task performance—neither H123 nor Trace reaches competitive zero-shot accuracy. TAM and WTCM, despite their popularity in prior single-modal tasks, show poor alignment and low performance in our cross-modal setup. This is likely due to their use of fixed-size sliding windows, which obscure packet-level semantic anchors and disrupt alignment with logic-side representations.

\begin{figure*}[tp]
    \centering
    \includegraphics[width=\linewidth]{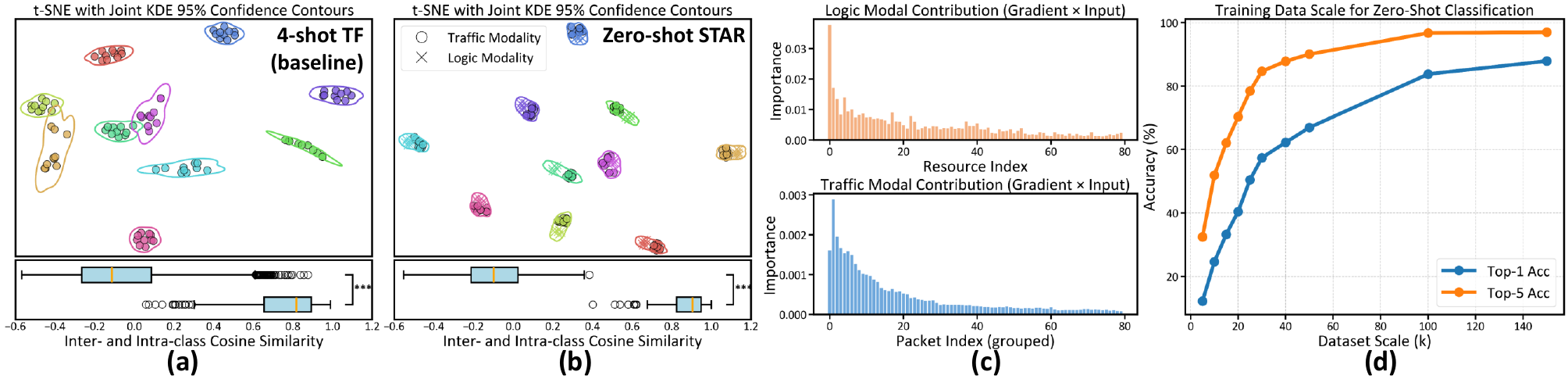}
    \caption{
        \textbf{Analysis of the STAR model.}  
(a, b) t-SNE visualization and cosine similarity statistics of modality representations learned by TF (baseline) and STAR, respectively.  
(c) Gradient-based importance scores over input positions reveal that both modalities exhibit localized discriminative patterns.  
(d) Impact of training data scale on zero-shot classification accuracy, showing rapid performance saturation after 100k samples.
    }
    \label{fig:last}
\end{figure*}

In contrast, our proposed traffic encoding preserves both protocol semantics and alignment structures, achieving balanced modality properties and significantly superior classification accuracy. These results validate our cross-modal formulation and modality design as crucial to enabling effective zero-shot retrieval.

\subsection{Closed-World Website Fingerprinting}
\label{subsec_cw}

To evaluate STAR under a standard closed-world setting, we conduct experiments where the client accesses a fixed set of monitored websites. Results are summarized in Table~\ref{tab:fewshot}.

\noindent\textbf{• Experimental Setup.} We train STAR on a mixture of three datasets: the STAR-200K cross-modal dataset, structure-aware augmented pairs and the labeled training portion of H\&W, combined at a 10:3:3 ratio. To assess zero-shot performance—i.e., recognizing websites unseen during training in the traffic modality—we construct disjoint training and evaluation website sets. The model is optimized with the objective in Eq.~\ref{eq_loss} for 200 epochs on 5 NVIDIA A100 GPUs, requiring about 4 hours. During inference, we follow a CLIP-style retrieval procedure: each traffic sample is embedded using the traffic encoder and projection head, and cosine similarity is computed against 1,600 logic-side anchors, each representing an embedding of a logic modality sample from the test set. Classification is determined by nearest-neighbor retrieval, and we report both top-1 and top-5 accuracy. 

Since no existing website fingerprinting approach supports zero-shot classification, we build a baseline using k-means clustering ($K=1,600$) with optimal label assignment obtained via the Hungarian algorithm. For few-shot evaluation, we follow the standard $n$-shot setting on H\&W-1600, using $n$ labeled samples per class. Competing methods (e.g., TF, NetCLR, H\&W) are trained on the $n$-shot subset, while STAR uses lightweight adaptation via a linear probe and Tip-Adapter.

\noindent\textbf{• Results.} STAR delivers strong \textit{zero-shot} performance, achieving \textbf{87.87\% top-1} and \textbf{96.94\% top-5 accuracy} over 1,600 website classes, despite not seeing any traffic samples from the evaluation set. This confirms the effectiveness and generalization ability of the learned cross-modal alignment. With few-shot adaptation, STAR’s performance improves further, reaching 95.06\% top-1 accuracy with a linear probe at 16-shot and up to 99.09\% with Tip-Adapter. Compared with existing few-shot methods, STAR provides both higher upper-bound accuracy and better zero-shot generalization. For example, H\&W and TF remain competitive under few-shot settings but still plateau below STAR’s adapted results. Notably, As shown in Fig. \ref{fig:cw_ow}(a), STAR’s zero-shot accuracy already matches the average 8-shot performance of other methods, which typically require over 100 hours of traffic collection on a single machine~\cite{tf}, highlighting its advantage in low-data and real-time deployment scenarios.

\subsection{Open-World Website Fingerprinting}
\label{subsec_ow}
Website fingerprinting in the open-world setting offers a more realistic evaluation paradigm, as it assumes the client may access websites outside the attacker's monitored set. In this scenario, effective attack methods must possess the capability to reject inputs from unknown or unmonitored websites—an open-set recognition challenge.

\noindent\textbf{• Experimental Setup.}
To evaluate performance under open-world conditions, we select the top-performing baselines from the closed-world experiments: CountMamba, DF+, and H\&W, representing state-of-the-art methods across different paradigms. Each is trained under a 4-shot setting (i.e., 4 labeled examples per monitored class), and compared against our \textit{zero-shot} \textbf{STAR} model.

These four methods span two representative open-world handling strategies:
\textbf{(i)} Threshold-based similarity rejection: STAR and H\&W directly compute similarity scores between test samples and support samples, rejecting a test input if its top similarity falls below a predefined threshold.
\textbf{(ii)} Explicit background class training: CountMamba and DF+ introduce a “non-monitored” class during training by sampling from a large unmonitored website set, enabling the model to learn a decision boundary between monitored and unmonitored traffic.

Following established practice \cite{hw}, we adopt a binary classification evaluation strategy between monitored and unmonitored samples. At test time, we construct a balanced evaluation set, ensuring a 1:1 ratio of monitored and unmonitored samples. We report precision and recall at varying decision thresholds, along with the overall AUC and the best F1 score for each method.

\noindent\textbf{• Results.}
Experimental results are summarized in Fig. \ref{fig:cw_ow}(b). \textit{zero-shot} STAR achieves the best open-world detection performance among all evaluated methods, attaining an \textbf{AUC of 0.963}, significantly outperforming CountMamba (0.926), DF+ (0.854), and H\&W (0.884). The corresponding best F1 score of STAR is 90.65, indicating both high precision and strong recall.

We attribute this performance gain to the \textbf{cross-modal alignment learning paradigm} employed by STAR. Unlike traditional classification-based approaches—which focus on optimizing decision boundaries over a fixed set of monitored classes—our model is trained on large-scale cross-modal sample pairs, allowing it to learn a more \textbf{generalizable alignment space} between website-level semantic features and encrypted traffic patterns. This alignment is not bound to specific class labels, but rather captures discriminative structure across the broader web domain. As a result, even in open-world settings where unseen websites appear, STAR can reliably identify whether a test sample aligns well with any monitored site—without requiring explicit negative class supervision during training.
These findings highlight the unique advantage of retrieval-based, modality-aligned approaches in realistic, open-set fingerprinting scenarios.

\begin{table}[t]
  \small
  \setlength{\tabcolsep}{7pt}  

  \begin{threeparttable}
    \caption{Ablation study results.}
    \label{tab:ablation}
    \begin{tabular*}{\columnwidth}{@{\extracolsep{\fill}} l | cc | cc}
      \toprule
      \multirow{2}{*}{\textbf{Method}} 
        & \multicolumn{2}{c|}{\textbf{Closed‑World}} 
        & \multicolumn{2}{c}{\textbf{Open‑World}} \\
      \cmidrule{2-3}\cmidrule{4-5}
        & \textbf{Top‑1} & \textbf{Top‑5}
        & \textbf{AUC}   & \textbf{F1}    \\
      \midrule
      Base\tnote{1}
        & 69.56 & 91.06 & 0.850 & 82.63 \\
      Base + CMA\tnote{2}
        & 80.31 & 92.91 & 0.897 & 85.91 \\
      Base + CMA + OT$_\mathrm{Cls}$\tnote{3}
        & 82.19 & 94.75 & 0.916 & 87.03 \\
      Base + CMA + OT$_\mathrm{Cons}$
        & 84.06 & 95.87 & 0.929 & 87.12 \\
      Base + CMA + OT$_\mathrm{Hybrid}$
        & \textbf{87.87} & \textbf{96.94} & \textbf{0.963} & \textbf{90.65} \\
      \bottomrule
    \end{tabular*}

    \begin{tablenotes}[flushleft]
      \footnotesize
      \item[1] Basic cross‑modal alignment trained with large‑scale sample pairs using the InfoNCE loss.
      \item[2] \textbf{CMA}: Cross‑Modal Augmentation.
      \item[3] \textbf{OT}: Optimization Targets (see §\ref{subsec_loss}).
    \end{tablenotes}
  \end{threeparttable}
\end{table}

\subsection{Ablation and Interpretability Analysis}
\label{subsec_ablation}
To better understand the design, performance gain, and behavior of STAR, we conduct an in-depth analysis covering its key components, learned representations, and the effect of training scale.


\noindent\textbf{• Ablation Study.} Table~\ref{tab:ablation} reports ablation results under closed- and open-world settings. Starting from the base cross-modal alignment model trained with InfoNCE loss, incorporating \textbf{Cross-Modal Augmentation (CMA)} consistently improves performance. Additional gains are achieved by introducing \textbf{optimization targets} for classification (OT$_\mathrm{Cls}$), consistency (OT$_\mathrm{Cons}$), and their hybrid form, with OT$_\mathrm{Hybrid}$ yielding the best overall results.


\noindent\textbf{• Representation Analysis.} We visualize learned embeddings using t-SNE in \textbf{Fig. \ref{fig:last}(a-b)}. Compared with the TF baseline, STAR produces tighter intra-class clusters and stronger alignment between traffic and logic embeddings. Cosine similarity distributions further confirm higher intra-class similarity and improved separability.

\noindent\textbf{• Attribution and Scale Analysis.}
\textit{Gradient$\times$Input} attribution \cite{grad_input} (\textbf{Fig.~\ref{fig:last}(c)}) reveals that STAR exploits localized discriminative cues in both modalities. In the logic modality, influential features are concentrated in early resource slots, often corresponding to primary page elements, while in the traffic modality, early packet groups contribute disproportionately to alignment. \textbf{Fig.~\ref{fig:last}(d)} further shows that zero-shot accuracy improves rapidly with training scale and saturates beyond approximately 100k samples, suggesting diminishing returns once sufficient cross-modal diversity is learned.

Overall, this analysis demonstrates that STAR's superior performance arises not from any single component, but from the joint effect of robust cross-modal pretraining, effective alignment optimization, and scale-driven generalization.

\section{Discussion}

\subsection{Scope and Limitations}
This work redefines website fingerprinting as a cross-modal retrieval problem and presents STAR as a first realization of this paradigm. Our evaluation is scoped to standard HTTPS browsing sessions and validates the approach under typical conditions. More complex settings—such as multi-tab access, cross-network variability, and alternative encryption tunnels like VPN or Tor—remain beyond the scope of this initial study. We also focus on Chrome-based traffic traces, given its prevalence in practice; generalization to other browsers like Firefox, Safari remains to be assessed. These scenarios introduce additional factors that may affect alignment robustness and are left for future exploration.

\subsection{Implications and Future Directions}
STAR demonstrates that semantic–traffic alignment enables scalable, zero-shot fingerprinting without target-side traffic, revealing structural leakage as a persistent privacy risk even under full encryption. Beyond facilitating low-overhead deployment, our formulation offers a lens to analyze semantic leakage and guide defense design. Potential countermeasures include perturbing resource structures or obfuscating alignment anchors via traffic shaping, though their effectiveness and associated bandwidth overhead remain open challenges. Future work may extend STAR to multi-page tracking, dynamic contexts, or hybrid inference tasks involving both structure and behavior.

\section{Conclusion}
We reframed website fingerprinting under HTTPS as a zero-shot cross-modal retrieval problem and introduced STAR, a dual-encoder system that aligns semantic resource logic with encrypted traffic. Trained on large-scale logic–traffic pairs with structure-aware augmentation, STAR achieves strong zero-shot classification without target-side traffic collection. It surpasses state-of-the-art baselines across closed- and open-world settings, highlighting semantic–traffic alignment as a new axis of vulnerability. We release our dataset and implementation to support future research in both attack and defense directions.

\section*{Acknowledgment}
We sincerely thank the anonymous reviewers for their valuable comments and suggestions, which helped improve the quality of this paper.

\clearpage

\bibliography{references}

\end{document}